# Statistical Analysis of Weighted Networks

Antoniou Ioannis, Tsompa Eleni

Mathematics Department, Aristotle University, Thessaloniki, 54124, Greece

iantonio@math.auth.gr , etsompa@math.auth.gr

**Abstract**

The purpose of this paper is to assess the statistical characterization of weighted networks in terms of the generalization of the relevant parameters, namely average path length, degree distribution and clustering coefficient. Although the degree distribution and the average path length admit straightforward generalizations, for the clustering coefficient several different definitions have been proposed in the literature. We examined the different definitions and identified the similarities and differences between them. In order to elucidate the significance of different definitions of the weighted clustering coefficient, we studied their dependence on the weights of the connections. For this purpose, we introduce the relative perturbation norm of the weights as an index to assess the weight distribution. This study revealed new interesting statistical regularities in terms of the relative perturbation norm useful for the statistical characterization of weighted graphs.

## 1. Introduction

Complex systems may also [1] emerge from a large number of interdependent and interacting elements. Networks have proven to be effective models of natural or man-made complex systems, where the elements are represented by the nodes and their interactions by the links. Typical well known examples include communication and transportation networks, social networks, biological networks [2, 3, 4, 5].

Although the statistical analysis of the underlying topological structure has been very fruitful [2, 3, 4, 5] it was limited due to the fact that in real networks the links may have different capacities or intensities or flows of information or strengths. For example, weighted links can be used for the Internet, to represent the amount of data exchanged between two hosts in the network. For scientific collaboration networks the weight depends on the number of coauthored papers between two authors. For airport networks, it's either the number of available seats on direct flight connections between airports i and j or the actual number of passengers that travel from airport i to j. For neural networks the weight is the number of junctions between neurons and for transportation networks it's the Euclidean distance between two destinations.

The diversification of the links is described in terms of weights on the links. Therefore, the statistical analysis has to be extended from graphs to weighted complex networks. If all links are of equal weight, the statistical parameters used for unweighted graphs are sufficient for the statistical characterization of the network. Therefore, the statistical parameters of the weighted graphs should reduce to the corresponding parameters of the conventional graphs if all weights are put equal to unity.

Complex graphs are characterized by three main statistical parameters, namely the degree distribution, the average path length and the clustering coefficient. We shall briefly mention the definitions for clarity and for a better understanding of the proposed extensions of these parameters for weighted graphs.

The structure of a network with N nodes is represented by a NxN binary matrix $A = \{a_{ij}\}$, known as adjacency matrix, whose element $a_{ij}$ equals 1, when there is a link joining node i to node j and 0 otherwise (i, j=1,2,…,N).

In the case of undirected networks with no loops, the adjacency matrix is symmetric ($a_{ij} = a_{ji}$) and all elements of the main diagonal equal 0 ($a_{ii} = 0$).

The degree $k_i$ of a node i is defined as the number of its neighbours, i.e. the number of links incident to node i:

$$k_i = \sum_{j \in \Pi(i)} a_{ij} \qquad (1)$$

where $a_{ij}$ the elements of the adjacency matrix A and $\Pi(i)$ the neighborhood of node i.

The **degree distribution** is the probability that some node has k connections to other nodes and it is usually described by a power law $P(k) \sim k^{-\gamma}$, with $2 \leq \gamma \leq 3$.

The **characteristic path length of a network** is defined as the average of the shortest path lengths between any two nodes:

$$L = \frac{2}{N(N-1)} \sum_{i,j} d_{ij} \qquad (2)$$

where $d_{ij}$ is the shortest path length between i and j, defined as the minimum number of links traversed to get from node i to node j.

In many real networks it is found that the existence of a link between nodes i and j and between nodes i and k enhances the probability that node j will also be connected to node k. This tendency of the neighbours of any node i to connect to each other, is called clustering and is quantified by the **clustering coefficient** $C_i$, which is the fraction of triangles in which node i participates, to the maximum possible number of such triangles:

$$C_i = \frac{n_i}{k_i(k_i-1)} = \frac{\sum_{j,k} a_{ij} a_{jk} a_{ki}}{k_i(k_i-1)}, \quad k_i \neq 0,1 \qquad (3)$$

where $\frac{1}{2} n_i = \frac{1}{2} \sum_{j,k} a_{ij} a_{jk} a_{ki}$ is the actual number of triangles in which node i participates i.e. the actual number of links between the neighbours of node i, and $k_i(k_i-1)/2$ is the maximum possible number of links, when the subgraph of neighbours of node i is completely connected.

The clustering coefficient $C_i$ equals 1, if node i is the center of a fully interconnected cluster and equals 0, if the neighbours of node i are not connected to each other.

In order to characterize the network as a whole, we usually consider the average clustering coefficient C over all the nodes. We may also consider the average clustering coefficient C(k) over the node degree k.

Studies of real complex networks have shown that their connection topology is neither completely random nor completely regular, but lies between these extreme cases. Many real networks share features of both extreme cases. For example, the short average path length, typical of random networks, comes along with large clustering coefficient, typical of regular lattices. The coexistence of these attributes defines a distinct class of networks, interpolating between regular lattices and random networks, known today as small world networks [3, 4, 5, 6]. Another class of networks emerges when the degree distribution is a power law (scale free) distribution, which signifies the presence of a non negligible number of highly connected nodes, known as hubs. These nodes, with very large degree k compared to the average degree <k>, are critical for the network's robustness and vulnerability. These networks are known today as scale free networks [2, 3, 4, 7].

The purpose of this paper is to assess the statistical characterization of weighted networks in terms of proper generalizations of the relevant parameters, namely average path length, degree distribution and clustering coefficient. After reviewing the definitions of the weighted average path length, weighted degree distribution and weighted clustering coefficient in section 2, we compare them in section 3. Although the degree distribution and the average path length admit straightforward generalizations, for the clustering coefficient several different definitions have been proposed. In order to elucidate the significance of different definitions of the weighted clustering coefficient, we studied their dependence on the weights of the connections in section 4, where we introduce the relative perturbation norm as an index to assess the weight distribution. This study revealed new interesting statistical regularities in terms of the relative perturbation norm useful for the statistical characterization of weighted graphs.

## 2. Statistical parameters of weighted networks

The weights of the links between nodes are described by a NxN matrix $W = \{w_{ij}\}$. The weight $w_{ij}$ is 0 if the nodes i and j are not linked. We will consider the case of symmetric positive weights ($w_{ij} = w_{ji} \geq 0$), with no loops ($w_{ii} = 0$).

In order to compare different networks or different kinds of weights, we usually normalize the weights in the interval [0,1], by dividing all weights by the maximum weight. The normalized weights are $\dfrac{w_{ij}}{\max(w_{ij})}$.

The statistical parameters for weighted networks are defined as follows.

The node degree $k_i = \sum_{j \in \Pi(i)} a_{ij}$, which is the number of links attached to node i, is extended directly to the **strength or weighted degree**, which is the sum of the weights of all links attached to node i:

$$s_i = \sum_{j \in \Pi(i)} w_{ij} \tag{4}$$

The strength of a node takes into account both the connectivity as well as the weights of the links.

The degree distribution is also extended for the weighted networks to the **strength distribution P(s),** which is the probability that some node's strength equals s. Recent studies indicate power law $P(s) \sim s^{-a}$ [8, 9, 10].

There are two different generalizations of the characteristic path length in the literature, applicable to transportation and communication networks. In the case of transportation networks the **weighted shortest path length** $d_{ij}$ between i and j, is defined as the smallest sum of the weights of the links throughout all possible paths from node i to node j [11, 12]:

$$d_{ij} = \min \sum_{i,j} w_{ij} \tag{5}$$

The weight describes physical distances and/or cost usually involved in transportation networks. The capacity/intensity/strength/efficiency of the connection is inversely proportional to the weight.

However, this definition is not suitable for communication networks, where the efficiency of the communication channel between two nodes is proportional to the weight. The shortest path length in case of communication networks is defined as:

$$d_{ij} = \min \sum_{i,j} \frac{1}{w_{ij}} \tag{6}$$

To our knowledge, the latter definition has been used by Latora and Marchiori [13, 14] for the definition of the "efficiency" of the network, as inversely proportional to the shortest path length $d_{ij}$.

The weighted characteristic path length for both cases is the average of all shortest path lengths and it is calculated by formula (2).

We found in the literature six proposals for the definitions of the **weighted clustering coefficient**, which we shall review.

- Zhang et. al. (2005) [15] definition:

$$C_{w,i}^Z = \frac{\sum_j \sum_k w_{ij} w_{jk} w_{ki}}{\left(\sum_j w_{ij}\right)^2 - \sum_j w_{ij}^2} \tag{7}$$

The weights in this definition are normalized. The idea of the generalization is the substitution of the elements of the adjacency matrix by the weights in the nominator of formula (3), as for the denominator the upper limit of the nominator is obtained in order to normalize the coefficient between 0 and 1. The definition originated from gene co-expression networks.

As shown by Kalna et. al. (2006) [16] an alternative formula that may apply for this definition is

$$C_{w,i}^K = \frac{\sum_j \sum_k w_{ij} w_{jk} w_{ki}}{\sum_j \sum_{k \neq j} w_{ij} w_{ik}}$$

- Lopez-Fernandez et. al. (2004) [17] definition:

$$C_{w,i}^L = \sum_{j,k \in \Pi(i)} \frac{w_{jk}}{k_i (k_i - 1)} \tag{8}$$

The weights in this definition are not normalized. The idea of the generalization is the substitution of the number of links that exist between the neighbours of node i in formula (3) by the weight of the link between the neighbours j and k. The definition originated from an affiliation network for committers (or modules) of free, open source software projects.

- Onnela et. al. (2005) [18] definition:

$$C_{w,i}^O = \frac{\sum_{j,k} \left( w_{ij} w_{jk} w_{ki} \right)^{\frac{1}{3}}}{k_i (k_i - 1)} \tag{9}$$

The weights in this definition are normalized. The quantity $I(g) = \left( w_{ij} w_{jk} w_{ki} \right)^{\frac{1}{3}}$ is called "intensity" of the triangle ijk. The concept for this generalization is to substitute the total number of the triangles in which node i participates, by the intensity of the triangle, which is geometric mean of the links' weights.

- Barrat et. al. (2004) [8] definition:

$$C_{w,i}^B = \frac{1}{s_i (k_i - 1)} \sum_{j,k} \frac{w_{ij} + w_{ik}}{2} a_{ij} a_{jk} a_{ki} \tag{10}$$

The weights in this definition are not normalized. The idea of the generalization is the substitution of the elements of the adjacency matrix in formula (3), by the average of the weights of the links between node i and its neighbours j and k with respect to normalization factor $s_i(k_i-1)$ which ensures that $0 \leq C_{w,i}^B \leq 1$. This definition was used for airport and scientific collaboration networks.

- Serrano et. al. (2006) [19] definition

$$C_{w,i}^S = \frac{\sum_j \sum_k w_{ij} w_{ik} a_{kj}}{s_i^2(1-Y_i)} \quad (11)$$

where $Y_i = \sum_j \left(\frac{w_{ij}}{s_i}\right)^2$ has been named "disparity".

The weights in this definition are not normalized. This formula is used for the generalization of the average clustering coefficient with degree k, which has a probabilistic interpretation just as the unweighted clustering coefficient.

- Holme et. al. [20] definition:

$$C_{w,i}^H = \frac{\sum_j \sum_k w_{ij} w_{jk} w_{ki}}{\max(w_{ij}) \sum_j \sum_{k \neq j} w_{ij} w_{ik}} \quad (12)$$

The only difference between formulas (7) and (12) is that (12) is divided by $\max(w_{ij})$. We shall not discuss this definition in the comparison because the essence of the comparison is already addressed by definition (7).

- Li et. al. (2005) [21] definition of the weighted clustering coefficient, is another version of the Lopez-Fernandez proposal (8).

**3. The relation between the different weighted clustering coefficients**

1. All definitions reduce to the clustering coefficient (3), when the weights $w_{ij}$ are replaced by the adjacency matrix elements.

2. All weighted clustering coefficients reduce to 0 when there are no links between the neighbours of node i, that is when $a_{jk} = w_{jk} = 0$.

3. In the other extreme, all weighted clustering coefficient take the value 1 when all neighbours of node i are connected to each other. Formulas (7) and (8) take the value 1 if the weights between the neighbours of the node i are 1, independently of the weights of the other links. Formula (9) takes the value 1, if and only if all the weights are equal to 1. Formulas (10) and (11) take the value 1 for all fully connected graphs, independently of all the weights.
   These calculations are presented in Appendix A.

4. We calculated the values of the weighted clustering coefficients of node i participating in a fully connected triangle. Formulas (7) and (8) take the value $w_{jk}$ of the weight of the link between neighbours j and k, of node i. Formula (9) becomes equal to the intensity of the triangle $C_{w,i}^{O} = (w_{ij} w_{jk} w_{ki})^{1/3}$ for all nodes of the triangle. Formulas (10) and (11) take the value 1 for all fully connected graphs, independently of all weights.
   These calculations are presented in Appendix B.

**4. The dependence of the weighted clustering coefficients on the weights**

In order to understand the meaning of the different proposals-definitions (7), (8), (9) (10) and (11) of the weighted clustering coefficient we shall examine their dependence on the weights, without alteration of the topology of the graph. We simply examine the values of these definitions for different distributions of weights, substituting the nonzero elements of the adjacency matrix A by weights normalized between 0 and 1.

A way to distinguish and compare different weight distributions over the same graph, is in terms of the relative perturbation norm $\frac{\|A-W\|}{\|A\|}$, which gives the percentage of the perturbation of the adjacency matrix introduced by the weights. For simplicity, we considered the $L_2$ norm.

We shall examine now the dependence of the weighted clustering coefficient with respect to the relative perturbation norms for several different weight distributions as well as for different graphs. We have examined many networks from 20 up to 300 nodes with different topologies that were generated by the networks software PAJEK

[22]. The weights examined are randomly generated numbers following uniform or normal distributions with several parameter values, so that the percentages of the perturbations scale from 0-90% increasing by 10% at each perturbation. All simulations gave rise to the same results, figs. 3 and 4, representing the typical trends of random and scale free networks, figs.1 and 2. It is remarkable to emphasize again that in all cases the same trends appear demonstrating a clear dependence on the relative perturbation norm only and no dependence on the values of weights on specific links.

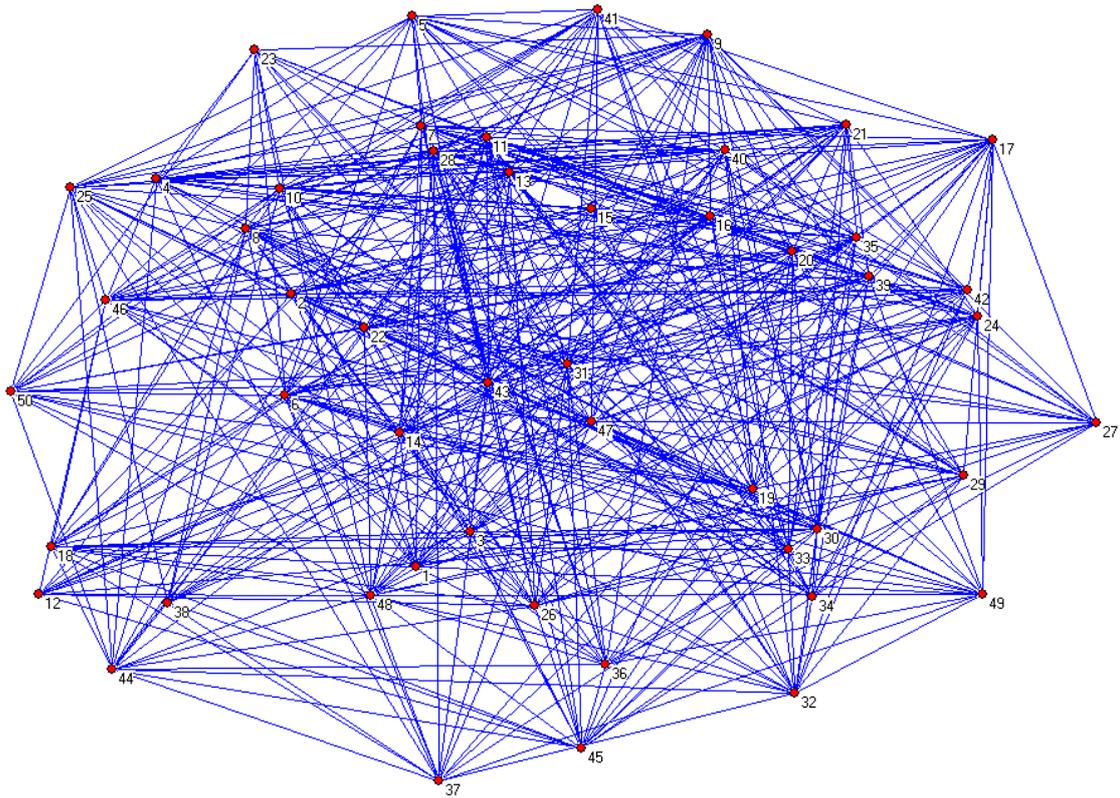

Figure 1. The random network (Erdos-Renyi model) examined consists of 100 nodes and was generated by the networks software PAJEK [22]. The clustering coefficient for the unweighted network is 0.3615.

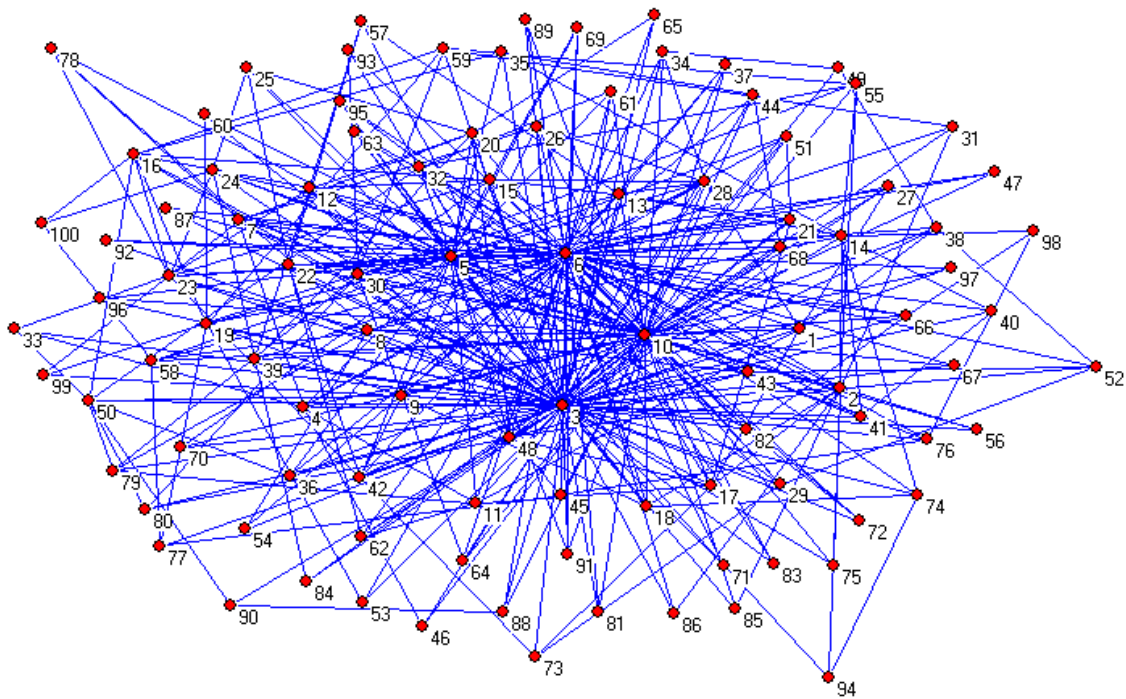

Figure 2. The scale-free network (Barabasi-Albert extended model) examined consists of 100 nodes and was generated by the networks software PAJEK [22]. The clustering coefficient for the unweighted network is 0.6561.

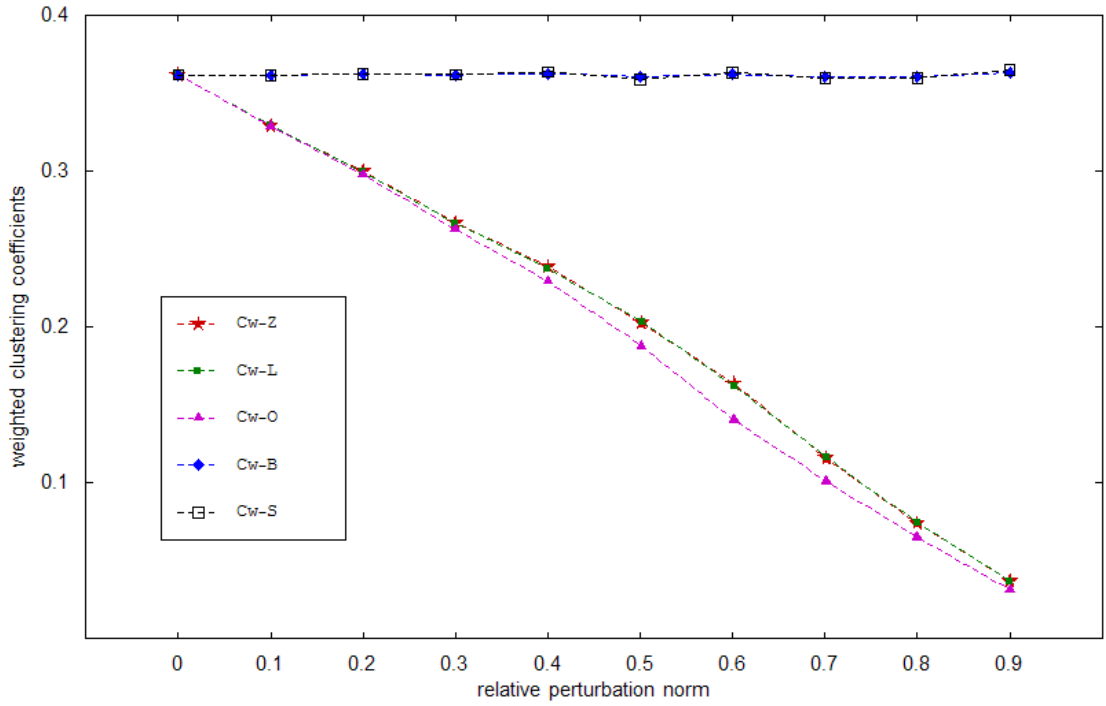

(A)

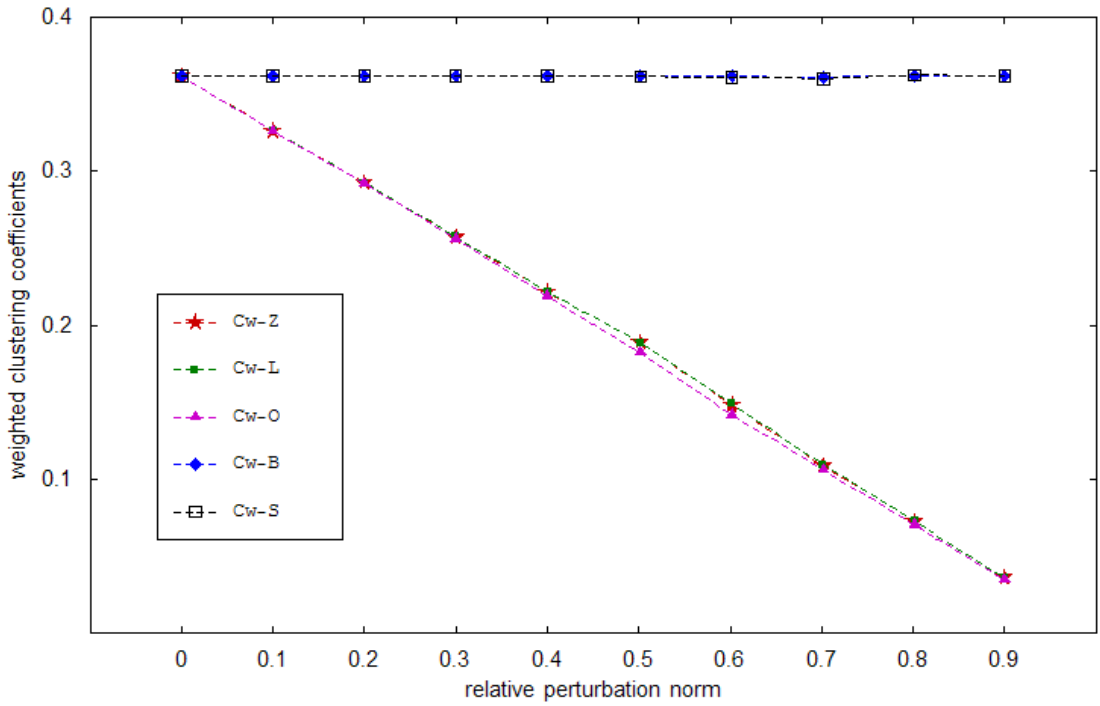

(B)

Figure 3. The values of all five weighted clustering coefficients Zhang et. al. $C_{w,i}^Z$ (★), Lopez-Fernandez et. al. $C_{w,i}^L$ (■), Onnela et. al. $C_{w,i}^O$ (▲), Barrat et. al. $C_{w,i}^B$ (◆) and Serrano et.al. $C_{w,i}^S$ (□), in terms of the relative perturbation norm for the random network (Erdos-Renyi model) with 100 nodes.

(A). The weights are randomly generated numbers following the uniform distribution.
(B). The weights are randomly generated numbers following the normal distribution.

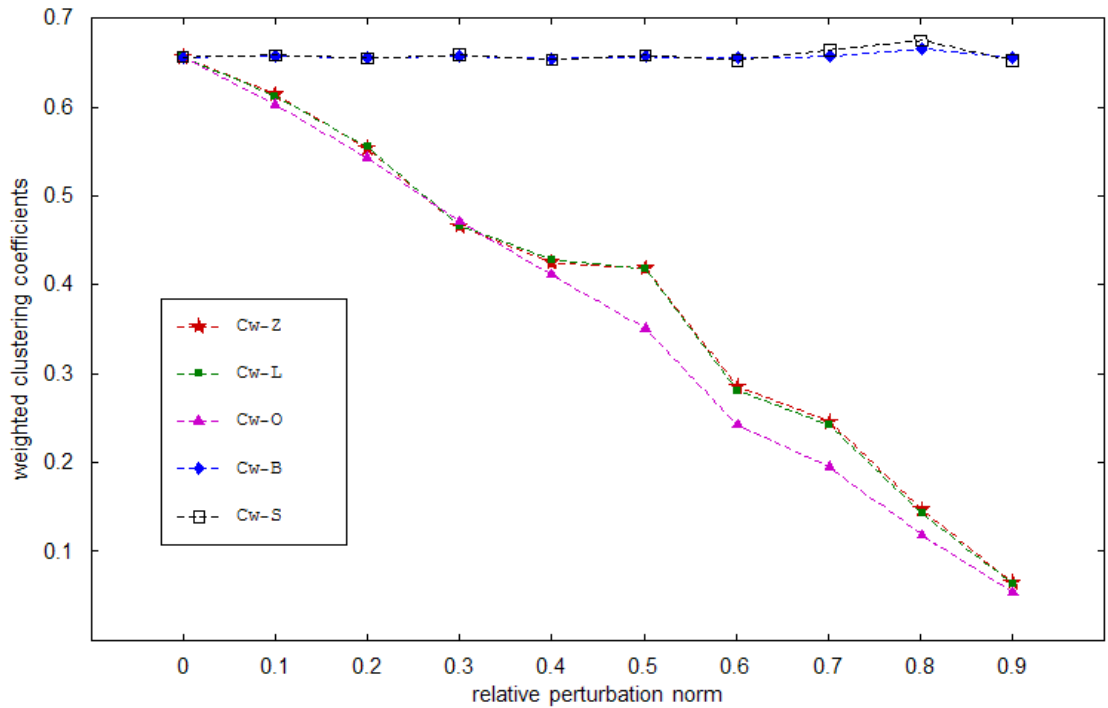

(A)

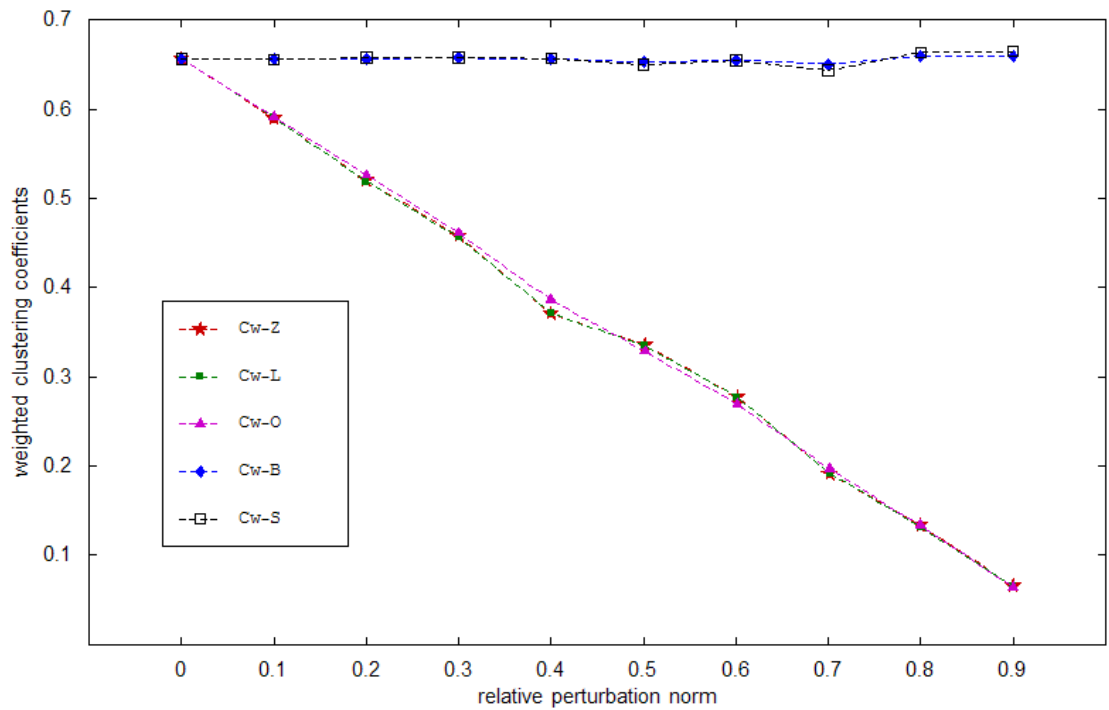

(B)

Figure 4. The values of all five weighted clustering coefficients Zhang et. al. $C_{w,i}^{Z}(\star)$, Lopez-Fernandez et. al. $C_{w,i}^{L}(\blacksquare)$, Onnela et. al. $C_{w,i}^{O}(\blacktriangle)$, Barrat et. al. $C_{w,i}^{B}(\blacklozenge)$ and Serrano et.al. $C_{w,i}^{S}(\square)$, in terms of the relative perturbation norm for the scale free network (Barabasi-Albert extended model) with 100 nodes.

(A). The weights are randomly generated numbers following the uniform distribution.
(B). The weights are randomly generated numbers following the normal distribution.

We observe in all cases a clear trend dependence of the values of all five weighted clustering coefficients, in terms of the relative perturbation norm of the weighted network. This demonstrates clearly first of all that the relative perturbation norm is a reliable index of the weight distribution. The Zhang et. al. (7), Lopez-Fernandez et. al. (8) and Onnela et. al. (9), weighted clustering coefficients follow the same trend, decaying smoothly as the relative perturbation norm increases. More specifically the trends of Zhang et. al. (7) and Lopez-Fernandez et. al. (8) almost coincide, while the trend of Onnela et. al. (10) varies slightly from the other two.

The weighted clustering coefficients of Barrat et. al. (10) and Serrano et.al. (11) do not change (variations appear after the first two decimal digits), regardless of the size of the network or the distribution of the weights. As mentioned in section 3, these coefficients are independent of the weights when the graph is completely connected. We notice here however, that weighted clustering coefficients (10) and (11) are independent of the weights for any graph.

**5. Concluding remarks**

1. The clear trend dependence of the values of all five weighted clustering coefficients in terms of the relative perturbation norm shows that the proposed relative perturbation norm is a reliable index of the weight distribution. The meaning of the decaying trend of weighted clustering coefficients Zhang et. al. (7), Lopez-Fernandez et. al. (8) and Onnela et. al. (9), with respect to the increase of the relative perturbation norm is quite natural. The clustering decreases almost linearly as the weights "decrease".
2. We presented in Appendices A and B the calculations demonstrating that all definitions reduce to the clustering coefficient (3), when the weights $w_{ij}$ are replaced by the adjacency matrix elements. The values of the weighted clustering coefficients of node i participating in a fully connected triangle are presented for completeness because we did not found them in the literature.
3. The results presented in figures 3 and 4 were necessary to obtain in order to have a minimal understanding of the statistical analysis of weighted networks, in order to proceed to applications on real networks.

**Acknowledgements**

We would like to thank Prof. Kandylis D. from the Medical School of Aristotle University of Thessaloniki who showed to us the significance of weighted networks in cognitive processes. We also thank Drs. Serrano M. A., Boguñá M. and Pastor-Satorras R. who pointed out their work to us.

## APPENDIX A. Calculations on the weighted clustering coefficient

The definitions (7)-(11) reduce to the clustering coefficient (3), when the weights $w_{ij}$ are replaced by the adjacency matrix elements.

1. Zhang et. al. (2005)

$$C_{w,i}^Z = \frac{\sum_j \sum_k w_{ij} w_{jk} w_{ki}}{\left(\sum_j w_{ij}\right)^2 - \sum_j w_{ij}^2}$$

The proof is presented by the authors.
For example, for a fully connected network with four nodes

$$C_{w,1}^Z = \frac{\sum_{j\neq 1}^4 \sum_{k\neq 1}^4 w_{1j} w_{jk} w_{k1}}{\left(\sum_{j\neq 1}^4 w_{1j}\right)^2 - \sum_{j\neq 1}^4 w_{1j}^2} = \frac{\sum_{j\neq 1}^4 w_{1j}\left(w_{j2}w_{21} + w_{j3}w_{31} + w_{j4}w_{41}\right)}{\left(w_{12} + w_{13} + w_{14}\right)^2 - w_{12}^2 - w_{13}^2 - w_{14}^2} =$$

$$= \frac{w_{12}w_{23}w_{31} + w_{12}w_{24}w_{41} + w_{13}w_{32}w_{21} + w_{13}w_{34}w_{41} + w_{14}w_{42}w_{21} + w_{14}w_{43}w_{31}}{w_{12}w_{13} + w_{13}w_{14} + w_{12}w_{14}} =$$

$$= \frac{w_{12}w_{23}w_{31} + w_{13}w_{34}w_{41} + w_{12}w_{24}w_{41}}{w_{12}w_{13} + w_{13}w_{14} + w_{12}w_{14}}$$

$C_{w,1}^Z = 1$ when $w_{23} = w_{34} = w_{24} = 1$

2. Lopez-Fernandez et. al. (2004)

$$C_{w,i}^L = \sum_{j,k \in \Pi(i)} \frac{w_{jk}}{k_i(k_i - 1)}$$

this formula can be expressed as

$$C_{w,i}^L = \frac{\sum_{j,k} w_{jk} a_{ij} a_{ik}}{k_i(k_i - 1)}$$

It is obvious that the formula reduces to the unweighted (3) when $w_{jk}$ are substituted by $a_{jk}$.

3. Onnela et. al. (2005)

$$C_{w,i}^O = \frac{\sum_{j,k}\left(w_{ij} w_{jk} w_{ki}\right)^{\frac{1}{3}}}{k_i(k_i - 1)}$$

reduces to the unweighted definition (3) when $w_{jk}$ are substituted by $a_{jk}$.

$$(a_{ij})^{\frac{1}{3}} = a_{ij}, \text{ hence } \left(w_{ij}w_{jk}w_{ki}\right)^{\frac{1}{3}} = \left(a_{ij}a_{jk}a_{ki}\right)^{\frac{1}{3}} = a_{ij}a_{jk}a_{ki}$$

$$C_{w,i}^{O} = \frac{\sum_{j}\sum_{k}\left(a_{ij}a_{jk}a_{ki}\right)^{\frac{1}{3}}}{k_i(k_i-1)} = \frac{\sum_{j}\sum_{k}a_{ij}a_{jk}a_{ki}}{k_i(k_i-1)}$$

4. Barrat et. al. (2004)

$$C_{w,i}^{B} = \frac{1}{s_i(k_i-1)}\sum_{j,k}\frac{w_{ij}+w_{ik}}{2}a_{ij}a_{jk}a_{ki}$$

reduces to the unweighted definition (3) when $w_{ij}$ and $w_{ik}$ are substituted by the adjacency matrix elements.

$$s_i = \sum_{j\in\Pi(i)} w_{ij} = \sum_{j\in\Pi(i)} a_{ij} = k_i \text{ and } a_{ij}^2 = a_{ij}.$$

$$C_{w,i}^{B} = \frac{1}{k_i(k_i-1)}\sum_{j,k}\frac{a_{ij}+a_{ik}}{2}a_{ij}a_{jk}a_{ki} = \frac{1}{k_i(k_i-1)}\sum_{j,k}\frac{a_{ij}a_{ij}a_{jk}a_{ki}+a_{ik}a_{ij}a_{jk}a_{ki}}{2} =$$

$$= \frac{1}{k_i(k_i-1)}\sum_{j,k}\frac{a_{ij}^2 a_{jk}a_{ki}+a_{ij}a_{jk}a_{ki}^2}{2} = \frac{1}{k_i(k_i-1)}\sum_{j,k}\frac{a_{ij}a_{jk}a_{ki}+a_{ij}a_{jk}a_{ki}}{2} =$$

$$= \frac{1}{k_i(k_i-1)}\sum_{j,k}a_{ij}a_{jk}a_{ki}$$

5. Serrano et. al. (2006) formula can be expressed as

$$C_{w,i}^{S} = \frac{\sum_{j}\sum_{k}w_{ij}w_{ik}a_{kj}}{s_i^2(1-Y_i)} = \frac{\sum_{j}\sum_{k}w_{ij}w_{ik}a_{kj}}{s_i^2\left(1-\sum_{j}\left(\frac{w_{ij}}{s_i}\right)^2\right)} = \frac{\sum_{j}\sum_{k}w_{ij}w_{ik}a_{kj}}{s_i^2\left(1-\frac{1}{s_i^2}\sum_{j}w_{ij}^2\right)} = \frac{\sum_{j}\sum_{k}w_{ij}w_{ik}a_{kj}}{s_i^2-\sum_{j}w_{ij}^2} =$$

$$= \frac{\sum_{j}\sum_{k}w_{ij}w_{ik}a_{kj}}{\left(\sum_{j}w_{ij}\right)^2-\sum_{j}w_{ij}^2}$$

It is obvious that the formula reduces to the unweighted (3) when $w_{jk}$ are substituted by $a_{jk}$.

**APPENDIX B. The values of the weighted clustering coefficients of some node i participating in a fully connected triangle.**

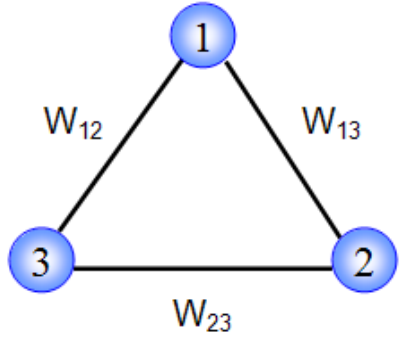

We calculate the weighted clustering coefficient of node 1.

1. Zhang et. al. (2005)

$$C_{w,1}^{Z} = \frac{\sum_{j\neq 1}^{3}\sum_{k\neq 1}^{3} w_{1j} w_{jk} w_{k1}}{\left(\sum_{j\neq 1}^{3} w_{1j}\right)^2 - \sum_{j\neq 1}^{3} w_{1j}^2} = \frac{\sum_{j\neq 1}^{3} w_{1j}\left(w_{j2} w_{21} + w_{j3} w_{31}\right)}{\left(w_{12} + w_{13}\right)^2 - w_{12}^2 - w_{13}^2} =$$

$$= \frac{w_{12} w_{23} w_{31} + w_{13} w_{32} w_{21}}{2 w_{12} w_{13}} = \frac{2 w_{12} w_{23} w_{31}}{2 w_{12} w_{13}} = w_{23}$$

2. Lopez-Fernandez et. al. (2004)

$$C_{w,1}^{L} = \frac{\sum_{k\neq 1}^{3}\sum_{j\neq 1}^{3} w_{jk}}{k_1(k_1 - 1)} = \frac{\sum_{k\neq 1}^{3}\left(w_{2k} + w_{3k}\right)}{2(2-1)} = \frac{w_{23} + w_{32}}{2} = w_{23}$$

3. Onnela et. al. (2005)

$$C_{w,1}^{O} = \frac{\sum_{j}\sum_{k}\left(w_{1j} w_{jk} w_{k1}\right)^{\frac{1}{3}}}{k_1(k_1 - 1)} = \frac{\sum_{j}\left(w_{1j}\right)^{\frac{1}{3}}\sum_{k}\left(w_{jk} w_{k1}\right)^{\frac{1}{3}}}{2(2-1)} =$$

$$= \frac{1}{2}\sum_{j}\left(w_{1j}\right)^{\frac{1}{3}}\left(\left(w_{j2} w_{21}\right)^{\frac{1}{3}} + \left(w_{j3} w_{31}\right)^{\frac{1}{3}}\right) =$$

$$= \frac{1}{2}\left[\left(w_{12}\right)^{\frac{1}{3}}\left(w_{23} w_{31}\right)^{\frac{1}{3}} + \left(w_{13}\right)^{\frac{1}{3}}\left(w_{32} w_{21}\right)^{\frac{1}{3}}\right] =$$

$$= \frac{1}{2}\left[\left(w_{12} w_{23} w_{31}\right)^{\frac{1}{3}} + \left(w_{13} w_{32} w_{21}\right)^{\frac{1}{3}}\right] = \left(w_{12} w_{23} w_{31}\right)^{\frac{1}{3}} \leq 1$$

$$C_{w,1}^{O} = C_{w,2}^{O} = C_{w,3}^{O} = \left(w_{12} w_{23} w_{31}\right)^{\frac{1}{3}} \leq 1$$

4. Barrat et. al. (2004)

$$C_{w,i}^{B} = \frac{1}{s_i(k_i-1)} \sum_{j,k} \frac{w_{ij}+w_{ik}}{2} a_{ij}a_{jk}a_{ki}$$

Degree of node 1: $k_1 = \sum_{j \in \Pi(1)} a_{1j} = 2$

Strength of node 1: $s_1 = \sum_{j \in \Pi(1)} w_{1j} = w_{12}+w_{13}$

$$C_{w,1}^{B} = \frac{1}{s_1(k_1-1)} \sum_{j,k} \frac{w_{1j}+w_{1k}}{2} a_{1j}a_{jk}a_{k1} =$$

$$= \frac{1}{s_1(2-1)} \sum_{j} \left( \frac{w_{1j}+w_{12}}{2} a_{1j}a_{j2}a_{21} + \frac{w_{1j}+w_{13}}{2} a_{1j}a_{j3}a_{31} \right) =$$

$$= \frac{1}{s_1} \left( \frac{w_{13}+w_{12}}{2} a_{13}a_{32}a_{21} + \frac{w_{12}+w_{13}}{2} a_{12}a_{23}a_{31} \right) =$$

$$= \frac{1}{w_{12}+w_{13}} (w_{12}+w_{13}) a_{12}a_{23}a_{31} = a_{12}a_{23}a_{31} = 1$$

since $a_{12} = a_{23} = a_{31} = 1$

We also prove that Barrat et. al. definition for the weighted clustering coefficient is independent of all weights for all fully connected networks.

$$C_{w,i}^{B} = \frac{1}{s_i(k_i-1)} \sum_{j,h} \frac{w_{ij}+w_{ih}}{2} a_{ij}a_{jh}a_{hi} = \frac{1}{s_i(k_i-1)} \sum_{j}^{k_i} \sum_{h}^{k_i} \frac{w_{ij}+w_{ih}}{2} a_{ij}a_{jh}a_{hi} =$$

$$= \frac{1}{s_i(k_i-1)} \sum_{h}^{k_i} \left( \frac{w_{i1}+w_{ih}}{2} a_{i1}a_{1h}a_{hi} + \frac{w_{i2}+w_{ih}}{2} a_{i2}a_{2h}a_{hi} + ... + \frac{w_{ik_i}+w_{ih}}{2} a_{ik_i}a_{k_ih}a_{hi} \right) =$$

$$= \frac{1}{s_i(k_i-1)} \left[ \left( \frac{w_{i1}+w_{i1}}{2} a_{i1}a_{11}a_{1i} + \frac{w_{i2}+w_{i1}}{2} a_{i2}a_{21}a_{1i} + ... + \frac{w_{ik_i}+w_{i1}}{2} a_{ik_i}a_{k_i1}a_{1i} \right) + \right.$$

$$+ \left( \frac{w_{i1}+w_{i2}}{2} a_{i1}a_{12}a_{2i} + \frac{w_{i2}+w_{i2}}{2} a_{i2}a_{22}a_{2i} + ... + \frac{w_{ik_i}+w_{i2}}{2} a_{ik_i}a_{k_i2}a_{2i} \right) + ... +$$

$$\left. + \left( \frac{w_{i1}+w_{ik_i}}{2} a_{i1}a_{1k_i}a_{k_ii} + \frac{w_{i2}+w_{ik_i}}{2} a_{i2}a_{2k_i}a_{k_ii} + ... + \frac{w_{ik_i}+w_{ik_i}}{2} a_{ik_i}a_{k_ik_i}a_{k_ii} \right) \right]$$

For a fully connected network: $a_{ij} = 1, \forall i, j = 1, 2, ..., k_i$ and $a_{ii} = 0$, so

$$C_{w,i}^B = \frac{1}{s_i(k_i-1)} \left[ \left( 0 + \frac{w_{i2}+w_{i1}}{2} + \ldots + \frac{w_{ik_i}+w_{i1}}{2} \right) + \right.$$

$$+ \left( \frac{w_{i1}+w_{i2}}{2} + 0 + \ldots + \frac{w_{ik_i}+w_{i2}}{2} \right) + \ldots +$$

$$\left. + \left( \frac{w_{i1}+w_{ik_i}}{2} + \frac{w_{i2}+w_{ik_i}}{2} + \ldots + 0 \right) \right] =$$

$$= \frac{1}{s_i(k_i-1)} \left[ \left( \frac{w_{i1}+w_{i2}+\ldots+w_{ik_i}}{2} + (k_i-2)\frac{w_{i1}}{2} \right) + \right.$$

$$+ \left( \frac{w_{i1}+w_{i2}+\ldots+w_{ik_i}}{2} + (k_i-2)\frac{w_{i2}}{2} \right) + \ldots +$$

$$\left. + \left( \frac{w_{i1}+w_{i2}+\ldots+w_{ik_i}}{2} + (k_i-2)\frac{w_{ik_i}}{2} \right) \right] =$$

$$= \frac{1}{s_i(k_i-1)} \left( k_i \frac{w_{i1}+w_{i2}+\ldots+w_{ik_i}}{2} + (k_i-2)\frac{w_{i1}+w_{i2}+\ldots+w_{ik_i}}{2} \right) =$$

$$= \frac{1}{s_i(k_i-1)} (2k_i-2) \frac{w_{i1}+w_{i2}+\ldots+w_{ik_i}}{2} = \frac{w_{i1}+w_{i2}+\ldots+w_{ik_i}}{s_i} = 1$$

since $s_i = w_{i1} + w_{i2} + \ldots + w_{ik_i}$

5. Serrano et. al. (2006)

$$C_{w,1}^S = \frac{\sum_{j\neq 1}^{3} \sum_{k\neq 1}^{3} w_{1j} a_{jk} w_{k1}}{\left(\sum_{j\neq 1}^{3} w_{1j}\right)^2 - \sum_{j\neq 1}^{3} w_{1j}^2} = \frac{\sum_{j\neq 1}^{3} w_{1j}(a_{j2}w_{21} + a_{j3}w_{31})}{(w_{12}+w_{13})^2 - w_{12}^2 - w_{13}^2} =$$

$$= \frac{w_{12}a_{23}w_{31} + w_{13}a_{32}w_{21}}{2w_{12}w_{13}} = \frac{2w_{12}a_{23}w_{31}}{2w_{12}w_{13}} = 1$$

# References

[1] Prigogine I., (1980). *From being to becoming*, Freeman, New York.

[2] Barabasi, A.-L. (2002). *Linked: The New Science of Networks*, Perseus, Cambridge, MA.

[3] Dorogovtsev S.N., Mendes J.F.F. (2002). Evolution of networks, *Advances in Physics* **51**, 1079.

[4] Newman, M.E.J. (2003). The structure and function of complex networks, *SIAM Review* **45**, 167-256.

[5] Watts, D.J. (2003). *Six Degrees: The Science of a Connected Age*, Norton, New York.

[6] Watts, D.J., Strogatz, S.H. (1998). Collective dynamics of 'small-world' networks, *Nature* **393**, 440–442.

[7] Barabasi, A.-L., Albert, R. (1999). Emergence of scaling in random networks, *Science* **286**, 509-512.

[8] Barrat, A., Barthelemy, M., Pastor-Satorras, R., Vespignani, A. (2004). The architecture of complex weighted networks, *Proceedings of the National Academy of Sciences (USA)* **101**, 3747- 3752.

[9] Barthelemy, M., Barrat, A., Pastor-Satorras, R., Vespignani, A. (2005). Characterization and modeling of weighted networks, *Physica A* **346,** 34–43.

[10] Meiss, M. Menczer, F. Vespignani, A. (2005). On the Lack of Typical Behavior in the Global Web Traffic Network, *Proceedings of the 14th International World Wide Web Conference*, Japan, ACM 1595930469/05/0005.

[11] Thadakamalla, H.P., Kumara, S.R.T., Albert R. (2006). Search in weighted complex networks, cond-mat/0511476v2.

[12] Xu, X.-J., Wu, Z.-X., Wang Y.-H. (2005). Properties of weighted complex networks, cond-mat/0504294v3.

[13] Latora, V., Marchiori, M. (2001). Efficient behavior of small-world networks, *Physical Review Letters* **87**, 198701.

[14] Latora, V., Marchiori, M. (2003). Economic small-world behavior in weighted networks, *The European Physical Journal B* **32**, 249-263.

[15] Zhang, B., Horvath, S. (2005). A general framework for weighted gene co-expression network analysis, *Statistical Applications in Genetics and Molecular Biology*, **4**.

[16] Kalna, G., Higham, D.J. (2006). Clustering coefficients for weighted networks, *University of Strathclyde Mathematics Research report* **3**.

[17] Lopez-Fernandez, L., Robles, G., Gonzalez-Barahona, J.M. (2004). Applying social network analysis to the information in CVS repositories. *In Proc. of the 1st Intl. Workshop on Mining Software Repositories (MSR2004)*, 101-105.

[18] Onnela, J.-P., Saramäki, J., Kertész, J., Kaski, K. (2005). Intensity and coherence of motifs in weighted complex networks, *Physical Review E,* **71** (6), 065103.

[19] Serrano M. A., Boguñá M., Pastor-Satorras R. (2006). Correlations in weighted networks, *Physical Review E* **74**, 055101 (R)

[20] Holme P., Park S.M., Kim B.J., Edling C.R. (2004). Korean university life in a network perspective: Dynamics of a large affiliation network, cond-mat/0411634v1.

[21] Li, M., Fan, Y., Chen, J., Gao, L., Di, Z., Wu J. (2005). Weighted networks of scientific communication: the measurement and topological role of weight, *Physica A* **350**, 643–656.

[22] Batagelj V., Mrvar A.: Pajek. http://vlado.fmf.uni-lj.si/pub/networks/pajek/